\documentstyle[pra,twocolumn,aps]{revtex}
\input{epsf}
\begin{document}
\draft
\twocolumn[\hsize\textwidth\columnwidth\hsize\csname
@twocolumnfalse\endcsname
\title{Delocalizing effect of the Hubbard repulsion for electrons \\
on a two-dimensional disordered lattice}
\author{Bhargavi Srinivasan$^{(a)}$, Giuliano Benenti$^{(a,b)}$, 
and Dima L. Shepelyansky$^{(a)}$} 
\address{$^{(a)}$Laboratoire de Physique Quantique, UMR 5626 du CNRS,
Universit\'e Paul Sabatier, 31062 Toulouse Cedex 4, France}
\address{$^{(b)}$Center for Nonlinear and Complex 
Systems, Universit\`a degli Studi dell'Insubria and}
\address{Istituto Nazionale per la Fisica della Materia,
Unit\`a di Como, Via Valleggio 11, 22100 Como, Italy}  
\date{January  31, 2003}
\maketitle

\begin{abstract}
We study numerically  the ground-state properties of the repulsive 
Hubbard model
for spin-1/2 electrons on two-dimensional lattices with disordered
on-site energies. The projector quantum Monte Carlo method is 
used to obtain very accurate values of the ground-state charge
density distributions with $N_p$ and $N_p+1$ particles. The difference
in these charge densities allows us to study the localization properties
of an added particle. The results obtained at quarter-filling 
on finite clusters
show that the Hubbard repulsion has a strong delocalizing effect
on the electrons in disordered 2D lattices. However, numerical restrictions
do not allow us to reach a definite conclusion about the
existence of a metal-insulator transition in the thermodynamic limit
in two-dimensions.
\end{abstract} 
\vskip1pc]
\narrowtext

\section{Introduction}
The interplay between disorder and electron-electron interactions 
has been a subject of intense activity in the last few years. 
Recent experiments have shown the existence 
of an apparent metal-insulator transition (MIT) in 
two-dimensional (2D) semiconductor devices \cite{2DMIT}.  
This observation
came as a surprise to the community, since the scaling theory
of localization, developed for disordered, non-interacting systems
predicts insulating states even for infinitesimal disorder 
strengths \cite{lee}.
Therefore, these experiments have
promoted intense theoretical activity.
There is at present no consensus and considerable 
controversy surrounds this problem \cite{2DMIT}.
A novel feature of the experimental high mobility samples investigated is
their exceptionally low
electronic density $n_s$.
This leads to an unusually high value of the dimensionless
parameter $r_s \propto n_s^{-1/2}$ of up to 80.
The parameter $r_s$ sets the scale
of electron-electron interaction $E_{e-e}$ as compared to
the Fermi energy $E_F$ through
$r_s \approx E_{e-e}/E_F$. Thus the question 
of electron-electron interaction effects becomes important 
in these disordered systems. 

The analytical treatment of the problem of disordered,
interacting electrons is possible only in some limiting
cases. The effects of weak interactions in disordered 
systems have been studied in great detail in the metallic
(delocalized) regime \cite{altshuler}. The 
other extreme, corresponding to the strongly localized system
can be treated by mean-field methods \cite{efros}. Although
these analytical approaches have provided many
useful physical results, the general treatment of disordered
quantum many-body systems remains an unsolved problem.
Indeed, one of the most powerful methods developed for the 
analytical treatment of disordered systems, namely, the supersymmetry
approach \cite{supersym}, cannot handle the effects of electron-electron
interactions.  Given this context,
numerical approaches play a crucial role in
the treatment of disordered, interacting systems.

Several numerical approaches have been applied to the study
of 2D disordered, strongly correlated systems 
\cite{schreiber,benenti0,benenti,song,dassarma,pichard,benenti2,berkovits}. 
Among the approximate approaches, Hartree-Fock calculations with residual
interactions similar to the configuration interaction approaches
of quantum chemistry have been applied to systems
with spinless fermions and electrons with 
spin \cite{schreiber,benenti}. Exact diagonalization approaches
have been useful but suffer from  severe limitations in the
accessible system size and number of 
particles \cite{benenti0,dassarma,pichard,berkovits}. 
From these studies,
it has been observed that repulsive electron-electron interactions
can have a delocalizing effect in small systems. 
In addition, it has been shown experimentally \cite{2DMIT} 
that the spin degrees
of freedom play a crucial
role in the physics of these strongly interacting systems.
The inclusion of the spin degrees of freedom renders the 
numerical calculations even more difficult and strongly reduces
the number of fermions accessible 
(see e.g. \cite{schreiber,dassarma,pichard,benenti2,berkovits}).

Quantum Monte Carlo (QMC) approaches provide a powerful 
alternative to the treatment of quantum many-body systems.
These methods are in principle exact, apart from statistical
errors and allow the treatment of much larger system sizes
with many particles. The finite-temperature determinantal QMC
approach has been applied to the two-dimensional disordered 
Hubbard model and signatures of a metal-insulator transition
were obtained \cite{trivedi}. However, these calculations were
carried out at finite temperature and  could not access the ground
state of the system. In previous work, we studied the 
ground state properties of the disordered 2D Hubbard model
by the projector quantum Monte Carlo (PQMC) method \cite{caldara}.
We studied the properties of the Green's function, charge density
and inverse participation ratio against model parameters and
system size. While we observed some local charge reorganization,
we could not detect any significant delocalizing influence
of the Hubbard repulsion $U$ on the many-body ground state.
However, it should be noted that  all the physical quantities studied 
in Ref. \cite{caldara} were obtained by integrating
over all particles and effectively, the entire energy spectrum.
Here we introduce a new approach which allows us to
study the properties of a single added particle and therefore
emphasizes the physical effects of interactions in the proximity
of the Fermi edge.

This approach uses the inverse participation ratio (IPR) extracted
from the charge density differences of two many-body ground states,
as described below.
The IPR, $\xi$, for a normalized 
single-particle wavefunction
$\psi(i)$ is given by
$\xi = [\sum_i |\psi(i)|^4]^{-1}$, where $i$ is the site index of the system.
Clearly, $|\psi(i)|^2$ can be 
identified as the one-particle charge density at the site $i$.
This definition is usually
carried over to many-body systems by renormalizing the
total charge density at site $i$,  $\rho(i,N_p)$, 
which is obtained in the standard way from the ground 
state many-body wavefunction for $N_p$ particles. 
The IPR $\xi$ is then calculated
with the effective charge density $\rho(i,N_p)/N_p$.
We found that the IPR obtained from such a definition 
showed slight variations with model parameters and practically
no variation with system size \cite{caldara}. We believe that the
physical reason for this weak variance is the fact that this
procedure effectively integrates over all energies and therefore 
the dominant contribution comes from the states deeply below the 
Fermi energy, which remain strongly localized even in the presence of
interactions. Therefore, it is necessary to find a method which is more
sensitive to the contribution of states in the vicinity of the
Fermi energy. Recently, we studied the
localization properties of the disordered, {\it attractive}
Hubbard model in two- and three-dimensions \cite{attractive}.
In Ref. \cite{attractive}, we showed that an IPR calculated for
an added pair of particles is a very sensitive and relevant
quantity to study the localization properties of the
wavefunction.
Based on our results for the attractive Hubbard model, 
we have introduced a related quantity in the repulsive case, 
the IPR for a single added particle, which turns out to be much more 
sensitive to the effects of interaction. 
This quantity has certain similarities to the 
single-particle tunneling amplitude. The latter has recently been
studied by  exact diagonalization methods for small clusters with
spin-1/2 fermions \cite{berkovits}.
This work provided indications for a delocalizing effect induced
by repulsive interactions in disordered systems.
However, the number of particles studied was rather restricted due 
to the limitations of the exact diagonalization methods.
With our method and extensive, highly accurate PQMC simulations, 
we study considerably larger numbers of particles
in presence of strong electron-electron interactions.
In this study we report a significant 
delocalizing influence of the Hubbard repulsion on 2D
disordered electronic systems.
This paper is organized as follows :
after this Introduction, we describe the method used and the tests 
performed in the next section. Our results and discussion are presented in detail
in the third section.

\section{Model and method}
The Hamiltonian studied in this paper is the disordered, repulsive 
Hubbard model given by:
\begin{equation}
\label{hamiltonian}
\begin{array}{l}
H = H_A + H_I  \\=
\Bigl(
 -t\sum\limits_{\langle ij \rangle, \sigma} 
 c^{\dagger}_{i,\sigma} c_{j,\sigma}
 + \sum\limits_{i,\sigma} 
 \epsilon_i c^{\dagger}_{i,\sigma} c_{i,\sigma} 
 \Bigr)
  +  U \sum\limits_{i} n_{i\uparrow}n_{i\downarrow} 
  \end{array}
  \end{equation}
where $c_{i\sigma}^\dagger$ ($c_{i\sigma}$)
creates (destroys) an electron at site $i$
with spin $\sigma$ and
$n_{ i\sigma}=c_{i\sigma}^{\dagger}
c_{ i\sigma}$ is the corresponding
occupation number operator.
The hopping term $t$ between nearest neighbor lattice sites
characterizes the kinetic energy and
the random site energies $\epsilon_{i}$ are taken from a box
distribution over $[-W/2,W/2]$.
The parameter $U$ measures the strength of the screened, 
repulsive Hubbard interaction ($U>0$).
We have considered both the one- and two-dimensional cases, with
periodic boundary conditions in all directions. In the 2D case, the
sites $i$  lie on a rectangular lattice of linear dimension $N_x,N_y$.
The system size $N = N_x \times N_y$ then follows accordingly 
in one- and two-dimensions.
In the limit $U$ = 0, this Hamiltonian reduces to the Anderson
model (given by $H_A$) which is a standard model for the study 
of disordered systems \cite{kramer}.
In the  absence of disorder, $W$ = 0, this Hamiltonian reduces to the 
usual Hubbard model, which is one of the best-studied models for
correlated electronic systems \cite{dagotto}.

We have studied this Hamiltonian by the projector 
quantum Monte Carlo (PQMC)
method \cite{imada} and exact diagonalization calculations. 
The PQMC method was initially developed to study the ground state of 
the Hubbard model (clean limit of Eqn. (1)). The method can 
be generalized,
in principle quite simply, to incorporate disorder via the random
site energies. However, the actual implementation and convergence 
of the algorithm \cite{caldara} is highly non-trivial compared to the pure case,
 as will be discussed in detail below.

The PQMC method 
consists in filtering out the true ground state  $|\psi_{0} \rangle $ 
of the many-body system
from an appropriately chosen trial function $|\phi\rangle$:
\begin{equation}
|\psi_{0} \rangle
= \lim\limits_{\Theta \rightarrow \infty} { {e^{-
\Theta H}
|\phi\rangle} \over
 {\sqrt{\langle \phi| e^{- 2 \Theta H} |\phi \rangle}}}.
\label{projec}
\end{equation}
This method is exact in principle, apart 
from statistical errors and the sign problem  which appears for 
fermions at $U > 0$.
The Hamiltonian plays the role of the projection operator through the
term $e^{-  \Theta H}$, where $\Theta$ plays the role of the 
projection parameter.  The trial wave-function is usually formed
as a product of up and down spin states from the eigenstates of the 
non-interacting Hamiltonian. In our case, we chose the Fermi sea of
$H_A$ as the trial wave-function. In the PQMC procedure, 
the projection operator $\exp(-\Theta H)$
is first Trotter decomposed as
$\Bigl( \exp(- \Delta \tau H_{A})
 \exp(- \Delta \tau H_{I})  {\Bigr)}^{L}$ with
 $\Theta = \Delta\tau \times L$. This introduces a systematic error
 of order $(\Delta \tau)^2$ due to non-commutation of
 $H_A$ and $H_I$. We have used the symmetric Trotter 
 decomposition, which introduces a systematic error of
  $(\Delta \tau)^3$. The interaction is then decoupled by
 a discrete Hubbard-Stratonovich transformation, by the
 introduction of $N \times L$ Ising-like fields. This 
 Ising  model with complicated effective interactions is then treated
 by a Monte Carlo (MC) procedure to obtain the ground state 
 properties of the
 system.  The quantity calculated during the simulation is the 
 zero-temperature, equal-time Green's function, 
 $G_{ij} =\sum_{\sigma} \langle \psi_0 |
 c^{\dagger}_{i,\sigma} c_{j,\sigma}| \psi_0 \rangle$. 
 During the simulation this Green's 
 function can  be
 used to obtain all the other static correlation functions. 
 In the  algorithm used $O(N^2)$ operations are required to update the 
 Green's function
 after a MC step.  This procedure introduces cumulative errors and 
 therefore the Green's function has to be recalculated from scratch 
 regularly during the simulation (every $L_C$ steps), which requires
 $O(N^3)$ operations. 
 When the projection parameter $\Theta$
 becomes large, which is necessary for good convergence, the different
 components of the wavefunction tend to become parallel during the 
 projection process. Therefore, it is necessary to reorthogonalise the 
 components of the wavefunction regularly, every $L_R$ time steps. 
 In addition, it is well known that 
 quantum simulations of fermionic systems suffer from the sign
 problem except in some special cases. However, it has been observed
 that disorder in fact diminishes the magnitude of the sign problem.
 In our simulations, the sign problem is well under control, with
 the number of negative signs being less than $1\%$ of the total number
 of steps considered.

 We have studied systems  with particle number 
 ($N_p$) up to 25 fermions on lattice
 sizes of up to $6\times8$ sites,  with particle density always around
 quarter-filling.
 The simulations were carried out
 in the $S_z = 0$ sector for even numbers of particles and the 
 $S_z = 1/2$ sector for odd number of particles.
 The specific quantity
 studied in this paper is the difference of charge densities $\delta\rho(i)$
 between systems with $N_p$ and $N_p+1$ particles (see the next section for
 more details).  Here, we discuss the convergence of the PQMC
calculations.  The charge densities involved in
the calculation $\delta\rho(i)$ are 
obtained from two independent simulations of the same disorder
realization with different particle numbers, $N_p$ and $N_p +1$.
 Therefore, it becomes necessary
 to measure very accurately the distribution of this added particle
 over the lattice. Clearly, it  is harder to measure $\delta\rho(i)$ accurately
 than measuring the total charge densities $\rho(i,N_p)$ in the ground state
 with $N_p$ particles. We carried out extensive tests
to verify the quality of our  $\delta\rho(i)$ data,  by
 varying the  PQMC parameters until convergence was obtained. We have
tested for convergence by varying the following PQMC parameters: the 
projection parameter $\Theta$,
 the Trotter time-step $\Delta\tau$, the Monte Carlo  parameters, 
 the reorthogonalization
 interval $L_R$ and the interval to recalculate the Green's function
 $L_C$. We have gone up to $\Theta = 15$ and $\Delta\tau = 0.05$. For
the physical parameters used in this paper, we find that $\Theta = 10$ with
$\Delta \tau = 0.1$ are sufficient. This corresponds to a systematic error
of $10^{-3}$. After varying the parameters $L_R$ and $L_C$ 
we  decided to recalculate the Green's function  with
reorthogonalization of the components, after every 5 time-steps
( $L_R = L_C = 5$).  We have also checked the  MC parameters and find that
 3000 sweeps are adequate for convergence, with 1000 sweeps for 
 equilibration.

The disorder average
was carried out over $N_R$ different disorder realizations with 
$N_R = 16$ in the PQMC
simulations and $N_R = 100$  for most exact calculations.
The site-energies are randomly chosen for $N_R$ disorder realizations at
$W/t = 1$ and then scaled proportionally to $W/t$ for stronger values of
disorder.

While our main motivation is the study of the 2D case, there
are many useful reasons to study 1D systems. The 1D case
can be studied conveniently by exact diagonalization methods, while
changing the system size. Thus, we can study the variation of
various properties such as the inverse participation ratio (IPR) $\xi$, 
with system size, exactly.  These calculations 
provide a strong independent check on  the PQMC data at small sizes. Further,
the localization effect is much stronger in 1D and $\delta\rho$ is usually
strongly peaked in 1D as compared to 2D.
Therefore, if the algorithm is capable of reproducing a localized
peak, this provides a strong check on the method, 
at the given range of physical parameters (examples provided 
in the next section).

\section{Results and Discussion}

In order to study the localization properties of the system,
we use the charge density distribution for an added particle
at the Fermi level, given by:
$\delta\rho(i)=\rho(i,N_p+1)-\rho(i,N_p)$,
where $\rho(i)$ is the ground state charge density 
at site $i$.
The values of $\rho(i,N_p)$, $\rho(i,N_p+1)$ 
are obtained from two independent PQMC simulations for the 
same disorder realization. At $U$ = 0, this $\delta\rho(i)$ is identically
equal to the one-particle probability distribution (added particle)
at the Fermi edge.  In the interacting case, this charge density
distribution ($\delta\rho(i)$)
is not generally equal to  a probability distribution. For example,
it is not necessarily positive definite. However, in all the cases
studied  we find that this quantity is always positive. Therefore,
$\delta\rho(i)$ can be considered to be similar to a one-particle 
probability distribution. Thus, we can associate an inverse participation 
ratio (IPR) for an added
particle, for a given disorder realization as:
$\xi=(\sum_{i}\delta\rho(i)^2)^{-1}$. The disorder averaged 
IPR is given by $ \langle \xi \rangle$. At $U=0$, the IPR is a
standard tool used to obtain the number of sites over which the 
particle is localized \cite{supersym}. 
 We have successfully used this approach in previous work
on the disordered, attractive Hubbard model for a quantitative description of
the localization properties of the ground state \cite{attractive}.

\begin{figure} 
\centerline{\epsfxsize=8.cm\epsffile{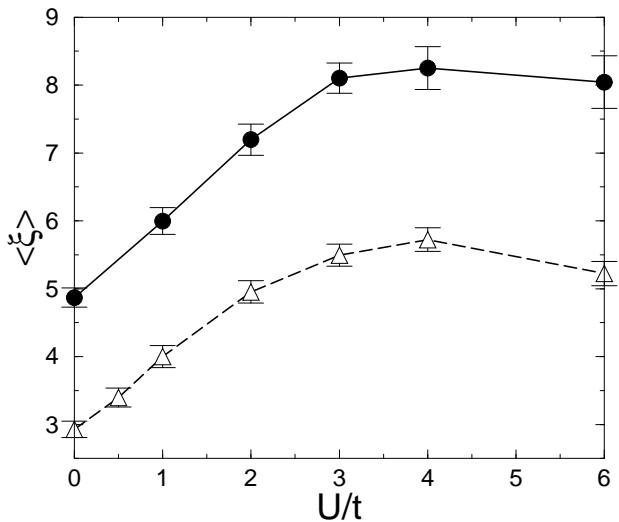}} 
\caption{IPR $\langle\xi\rangle$ versus 
interaction strength $U/t$, at $W/t=7$, for 
$N_p=6$ particles on a one-dimensional lattice 
with $N=12$ sites (triangles, average over 
100 disorder realizations), and on a $4\times 3$ 
lattice (circles, average over 16 disorder realizations).
Data comes from exact diagonalization. Here and in the 
following figures error bars denote statistical errors.}
\label{fig1} 
\end{figure} 

\begin{figure} 
\centerline{\epsfxsize=8.cm\epsffile{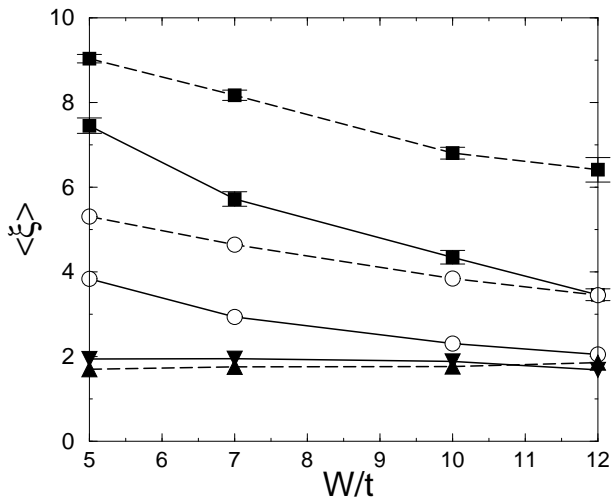}} 
\caption{IPR $\langle\xi\rangle$ versus 
disorder strength $W/t$. Continuous curves are
for the 1D system of 12 sites and $N_p=6$ particles, 
at $U/t=0$ (open circles) and $U/t=4$ (filled squares).
Curves with dashed lines are for the $4 \times 3$ lattice
and $N_p=6$ particles, 
at $U/t=0$ (open circles) and $U/t=4$ (filled squares). 
Data come from exact diagonalization and are averaged 
over 100 disorder realizations (except the point 
for the $4\times 3$ lattice at $U/t=4$, $W/t=12$,
with $N_R = 16$).
The continuous curve with inverted triangles is the ratio 
$\langle\xi(U/t)\rangle/\langle\xi(U/t =0)\rangle$,
at $U/t=4$ for 1D and the dashed curve with 
triangles the same ratio for 2D.}
\label{fig2} 
\end{figure} 

In Figs. 1-2, we study the behavior of the IPR as a function of
$U/t$ and $W/t$ to try and establish the interesting range of
physical parameters in this system.
In Fig. 1, we compare the IPR $\langle\xi\rangle$ 
for a 1D ring of 12 sites and a $4\times3$ lattice in 2D as a function
of interaction strength $U/t$. From the figure we see that increasing
$U/t$ from 0 tends to increase the IPR up to intermediate strengths 
of the interaction. Thus, the optimal value  appears to be 
around $U/t=4$ in both 1D and 2D. Indeed, for $U/t > 4$, the 
value of IPR starts to decrease.
In Fig. 2, we see the variation of $\langle\xi\rangle$ for
small system sizes for $5 \le W/t \le 12$. Since the ratio 
$\langle\xi(U/t = 4)\rangle/ \langle\xi(U/t = 0)\rangle$ is 
practically constant, this justifies
our choice of parameter range for $W/t$. Disorder strength $W/t < 5$ would lead 
to states with localization length larger than the accessible system
size even at $U/t =0$. Convergence of the PQMC calculations becomes
progressively more difficult for larger disorder strengths.
Therefore, we studied the variation of the 
IPR for $W/t  = 5, 7$ in the 2D case. 

\begin{figure} 
\centerline{\epsfxsize=8.cm\epsffile{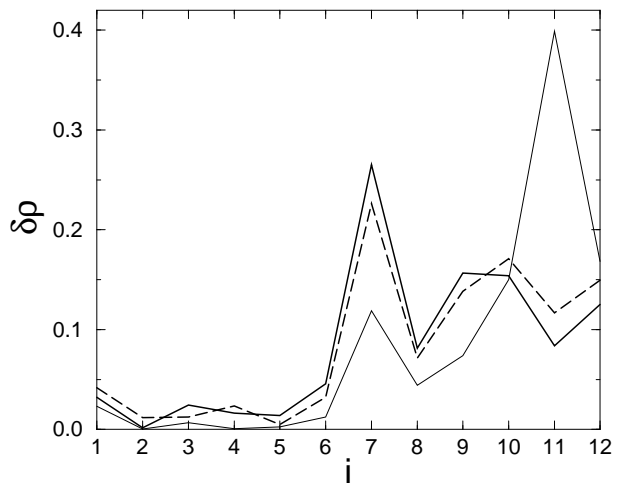}} 
\caption{Charge density difference $\delta\rho(i)$ for a 1D lattice
with $N=12$, $N_p=6$, $W/t=7$, $U/t=0$ (thin line, inverse 
participation ratio $\xi=4.31$), 
exact result for $U/t=2$ (thick solid line, $\xi=6.58$), and 
quantum Monte Carlo data for $U/t = 2$
(dashed line, $\xi=6.93$).}
\label{fig3} 
\end{figure}

Thus, it would seem ideal to study the IPR of 1D and 2D systems
for $5 \le W/t \le 7$ and $U/t \approx 4$. In 1D we could access 
these parameter values conveniently for the system sizes studied
by exact calculations. However, this becomes much more difficult
in 2D. This is because we study a small difference of large
total charge densities in the PQMC simulations. We found that the 
accuracy of our method for $\delta\rho(i)$ is good for 
for values of the interaction $U/t \leq 2$.
While we are not exactly at the optimal value of $U/t$, we are
nevertheless in a region with sufficiently strong interactions.
Indeed, it can be seen from Fig. 1 that $U/t = 2$ already has a
substantial delocalizing influence on the system. In Fig. 3, we 
present the PQMC calculation of $\delta\rho(i)$ in 1D at 
$U/t = 2$ as compared to exact calculations. We note that the
$\delta\rho(U=0)$ is strongly peaked. Increasing $U/t$ to 2
radically changes the picture and shifts the peak completely. 
The PQMC curve shown reproduces the quantitative picture of charge density
difference. It should
be noted that the calculation begins with  a trial wavefunction
corresponding to the $U/t =0$ data and changes completely to
give the correct physical picture. Quantitatively, the usual 
errors seen at these values of the physical parameters 
were around $3 - 5 \%$. The 2D case 
is more delocalized compared to 1D and we have observed that
convergence is better in 2D. 
For example, we have $\langle \xi \rangle$ = 7.19
(exact) and 7.43 (QMC) for a $4\times3$ lattice at $U/t = 2$ and
$W/t = 7$, averaged over 16 disorder realizations. This corresponds to
an error of about $3\%$ for the most extreme parameter values studied.
However, for $U/t = 4$, the error increases to $8 - 10\%$ and therefore
we restrict our studies to $U/t \leq 2$ in 2D.

\begin{figure} 
\centerline{\epsfxsize=8.cm\epsffile{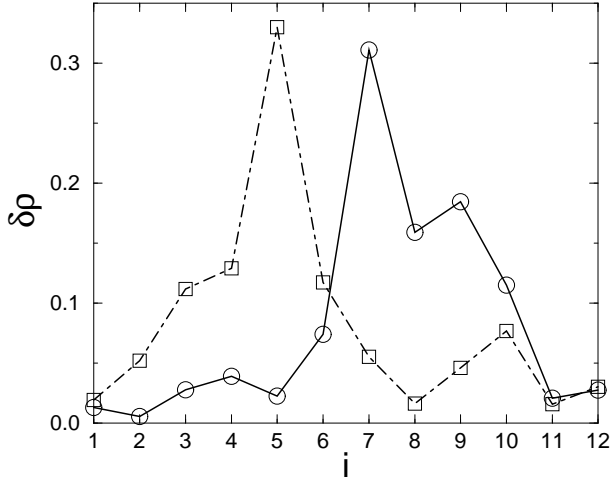}} 
\caption{Charge density difference $\delta\rho$ for a 1D lattice
with $N=12$, $N_p=6$, $W/t=7$, and $U/t=4$:
$\rho(N_p+1)-\rho(N_p)$ (continuous curve with circles) and 
$\rho(N_p+2)-\rho(N_p+1)$ (dot-dashed curve with squares).
Data are from exact diagonalizations, for the 
same disorder realization as in Fig. 3. For comparison, see data for
$\delta\rho$ at $U/t = 0$ in Fig. 3 (thin line).}
\label{fig4} 
\end{figure} 

Since we study a model of electrons with spin, it is
interesting to consider even-odd effects in the particle
number $N_p$ and system size $N$. In Fig. 4, we consider
the effect of progressively adding one and two particles 
to a 12 site ring with 6 electrons. From Fig. 3 and Fig. 4,
we see that the first added electron at $U/t$ = 2 and 4, 
has an entirely different peak as compared to the
non-interacting case. In Fig. 4, we see that the second
added electron also occupies a different region, as seen
from the position of the peak. This leads to the following physical 
image for the 1D case : each added electron finds its own 
location in space with optimal energy and it is well localized
by external disorder and random distribution of charges of other electrons.
This is in contrast with the non-interacting case, where the 
non-interacting orbital remains the same for odd and next even electron.

The situation is qualitatively different in 2D.
Indeed, in Fig. 5, we show  the
charge density difference for one and two added particles
on a $6 \times 6$ lattice with 18 electrons. This quantity
is obtained from PQMC simulations of the system for
a particular disorder realization, taken at $W/t$ = 7.
The data show that the initial $U/t = 0$ configuration
is very clearly localized for the given disorder realization
and lattice size. It is seen from the figure that the introduction
of a repulsive Hubbard interaction ($U/t = 2$) leads to 
a substantial delocalization of the added particle.
This is also borne out quantitatively, since the IPR
increases practically by a factor of 3, as compared 
to $U/t = 0$. In the non-interacting case, both added
particles occupy the same orbital. Therefore, the peak for
the second added particle, is trivially identical to the first.
It is remarkable that in the interacting case, the peak is
transformed to a much more extended distribution over the lattice.
Furthermore, the second added particle has 
practically the same distribution $\delta\rho(i)$ as the first. Thus, there 
appears to be almost zero effective repulsion between the two
added particles, despite the interaction strength $U/t = 2$.
This is completely in contrast to the scenario in 1D with interactions, 
where we observe significant repulsion between the two added 
particles. We note that this is the case for every disorder
realization studied. 

\begin{figure}
\centerline{\epsfxsize=4.2cm\epsffile{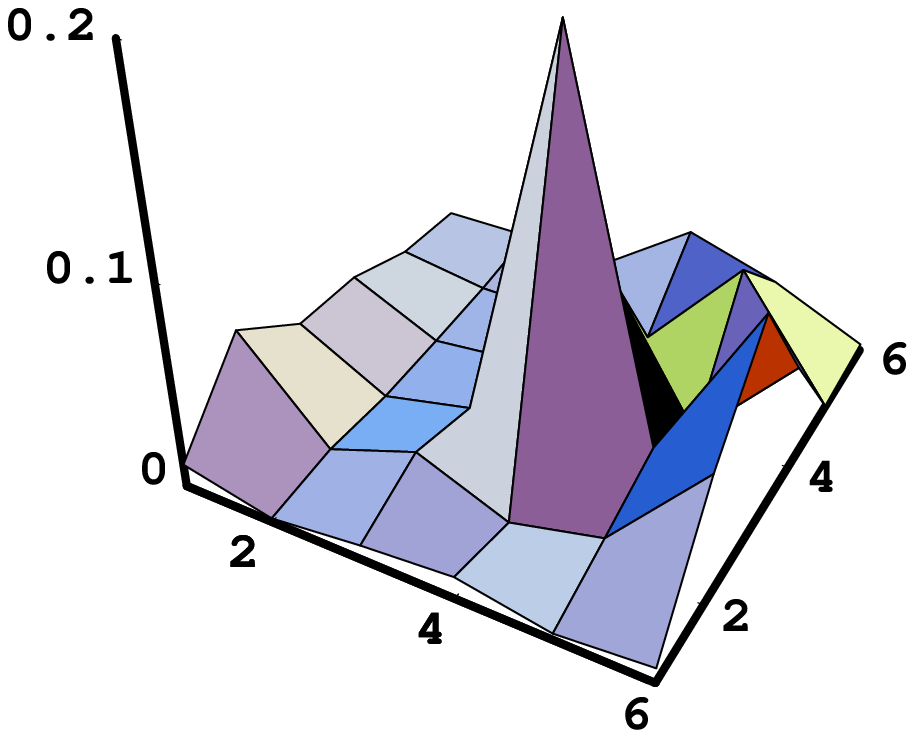}
\hfill\epsfxsize=4.2cm\epsffile{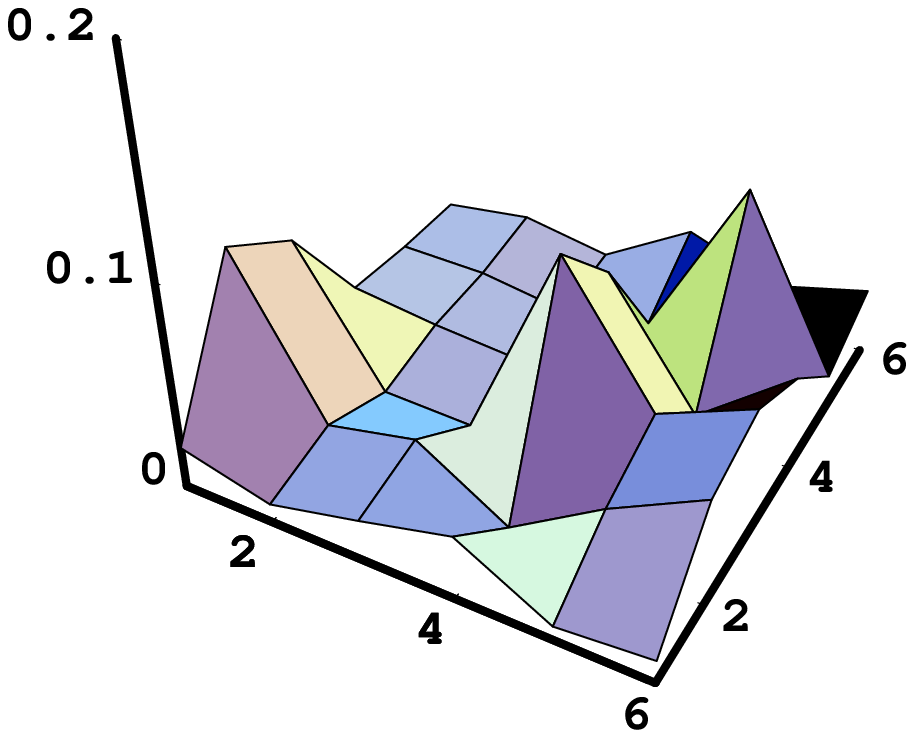}}
\vspace{-.5cm}
\centerline{\epsfxsize=4.2cm\epsffile{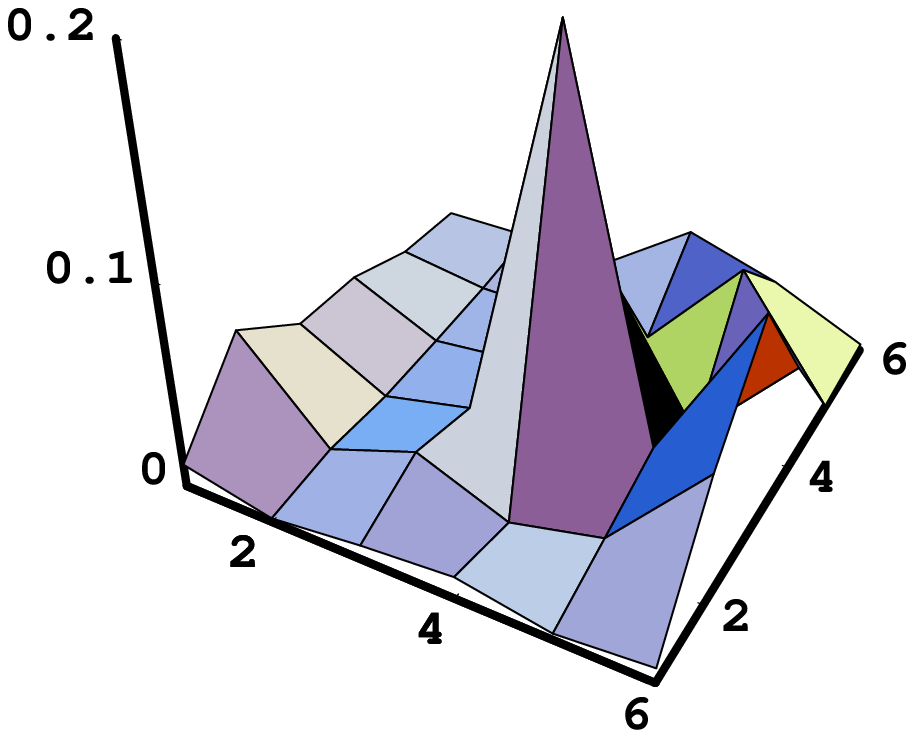}
\hfill\epsfxsize=4.2cm\epsffile{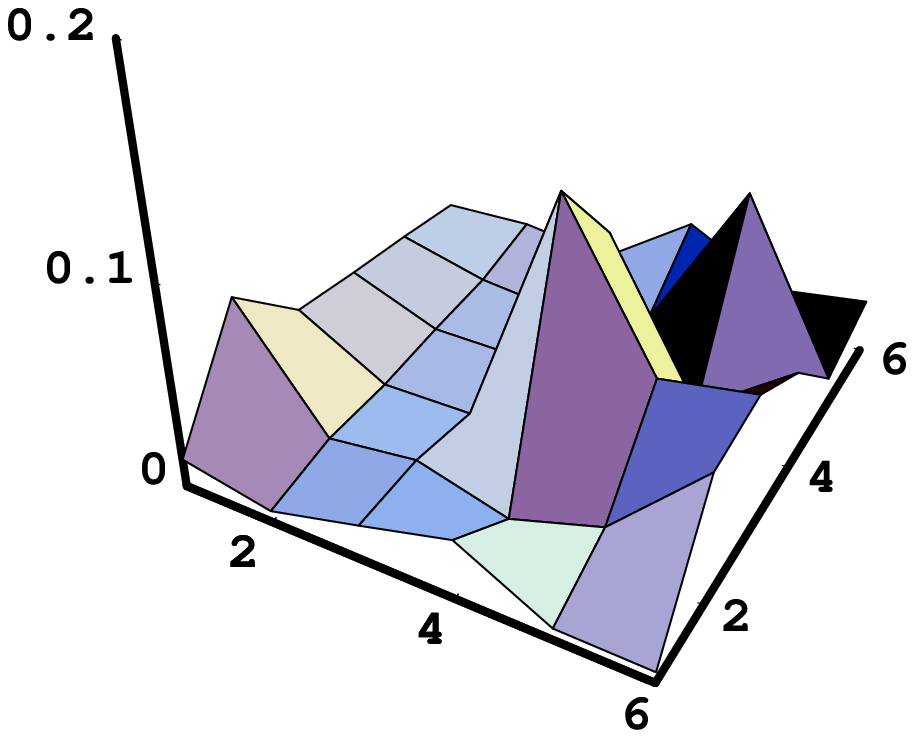}}
\caption{Charge density difference $\delta\rho$ for a 
$6\times 6$ lattice with $N_p=18$ particles,
$W/t=7$, $U/t=0$ (left) and $U/t=2$ (right): 
$\rho(N_p+1)-\rho(N_p)$ (top) and
$\rho(N_p+2)-\rho(N_p+1)$ (bottom).
Inverse participation ratios 
are $\xi=6.2$ (top left), $\xi=18.3$ (top right),
$\xi=6.2$ (bottom left), and $\xi=16.7$ (bottom right).}
\label{fig5}
\end{figure}

We now turn to a more quantitative picture of the one
and two dimensional systems. For this, it is necessary
to average the data over different disorder realizations
and to vary the system size. 
We start the analysis from the data for the 1D case.
In Fig. 6, we
present the IPR for interacting particles on 1D rings of 4--12 sites at 
constant density of particles (quarter filling)
and compare against the variation of the IPR at 
$U/t = 0$. We find a delocalization
effect even for 1D rings. We note that the curve for
$U/t = 0$ is well saturated in 1D even at 8--12 sites.
The $U/t = 2$ curve shows signs of saturation.
The delocalization effect is most visible for $U/t = 4$, as expected from
Fig. 1, where unfortunately, we do not have access to
PQMC data for comparisons. The ratio 
$\langle\xi(U/t)\rangle/\langle\xi(U/t =0)\rangle$
goes to 1.64 for $U/t = 2$ and 1.94 for $U/t = 4$.

\begin{figure} 
\centerline{\epsfxsize=8.cm\epsffile{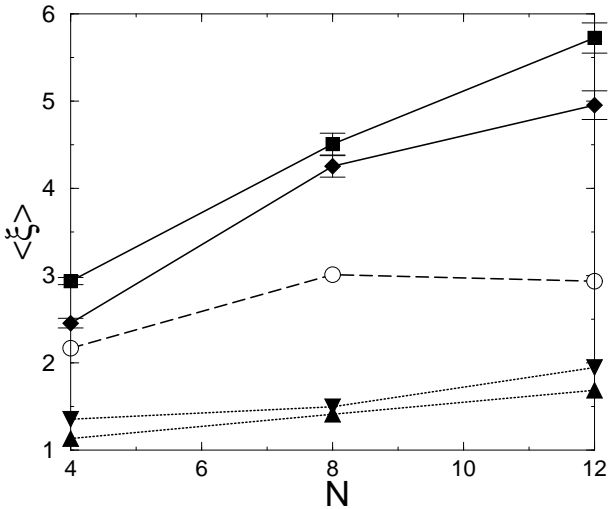}} 
\caption{Inverse participation ratio $\langle\xi\rangle$ versus 
system size $N$, for 1D chains with periodic boundary conditions,
at $W/t=7$, $U/t=0$ (circles), $U/t=2$ (diamonds), and
$U/t=4$ (squares).
Triangles and inverted triangles give the ratio 
$\langle\xi(U/t)\rangle/\langle\xi(U/t =0)\rangle$,
at $U/t=2$ and $U/t=4$, respectively.
Data come from exact diagonalization and are averaged over
100 disorder realizations.}
\label{fig6} 
\end{figure} 

In 2D previous exact diagonalization
studies have gone up to 6 electrons on a $4\times4$ lattice,
or 4 electrons on a $6\times6$ lattice 
\cite{dassarma,pichard,berkovits}.
Therefore, it is of great importance to access larger
system sizes with more particles.
In Fig. 7, we present the IPR for different lattice sizes
in 2D, obtained from PQMC simulations. 
We have gone up to
48 sites ($6 \times 8$ lattice) and 24-25 particles,
which goes beyond any existing study of the ground state 
in the literature.
We observe a  delocalizing effect of the Hubbard repulsion
manifested by an increase of the IPR with system size and interaction
strength.
There is a remarkable difference between the present result
and the IPR for the full many-body ground state wavefunction obtained in
Ref. \cite{caldara}.
In the previous work, we could notice
no size effects over a large range of $W$ and lattice sizes.
In fact, the curves for different lattice sizes against $U/t$ 
at a given value of $W/t$ tended to collapse as seen 
in Figs. 8 and 9 of Ref. \cite{caldara}. On the contrary, in the 
present work, we 
see from Fig. 7 that effects of lattice size on the IPR
for an added particle are considerable. 
Furthermore, we see  that the Hubbard repulsion has a strong delocalizing 
influence, compared to the data for the non-interacting case.  
As explained in the previous sections, we attribute
the difference between the present results and those presented
in Ref. \cite{caldara} to the fact that the IPR from $\delta\rho$ 
measures directly the response in the vicinity of the Fermi level.

\begin{figure} 
\centerline{\epsfxsize=8.cm\epsffile{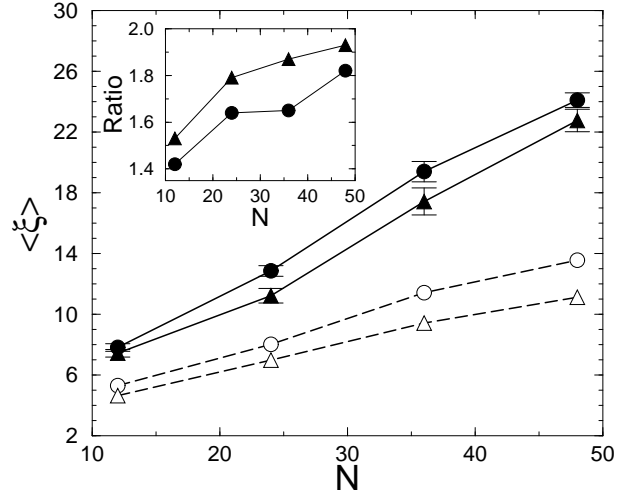}} 
\caption{Inverse participation ratio $\langle\xi\rangle$ as a function 
of the number of sites $N$, for a 2D lattice, with 
quarter filling
(i.e., $N_p=6,12,18$, and $24$ fermions on $4\times 3,
6\times 4,6\times 6$, and $8\times 6$ lattices,
respectively), with  $U/t=0$ (empty symbols, average over 
$10^3$ disorder realizations), $U/t=2$ (filled symbols,
average over $16$ disorder realizations), and $W/t=5$ (circles)
and $W/t=7$ (triangles). The inset shows the ratio 
$\langle\xi(U/t)\rangle/\langle\xi(U/t =0)\rangle$,
at $U/t=2$, for $W/t=5$ (circles) and 
$W/t=7$ (triangles). PQMC parameters: $\Theta=10$, 
$\Delta\tau=0.1$, except for $8\times 6$ 
lattices where $\Theta=12$, $\Delta\tau=0.08$.
}
\label{fig7} 
\end{figure}

The ratio
$\langle\xi(U/t)\rangle/\langle\xi(U/t =0)\rangle$
can be considered as a quantitative measure of 
delocalization effect induced by repulsive interaction.
From the inset of Fig. 7, it can be seen that this ratio
rises with system size for a given density and we do
not observe saturation at the system sizes studied.
Our data for this ratio for small clusters  in 2D are comparable
to the enhancement ratio obtained in Ref. \cite{berkovits} though
the quantity studied in Ref. \cite{berkovits} was slightly different i.e.
the one-particle tunneling amplitude. 
From Figs. 6 and 7 we can compare the values of the ratio 
$\langle\xi(U/t)\rangle/\langle\xi(U/t =0)\rangle$ in 1D and 2D.
At the system sizes studied these values are comparable for $U/t = 2$.
However, it would necessary to analyze the behavior of this ratio
for larger system sizes in order to get a clear answer to whether
the localization properties are indeed different in 1D and 2D.
At the same time, we note that there seem to be qualitative differences
based on even-odd effects for added particles in 1D and
2D (see Fig. 4 and Fig. 5 and discussion there). 
This difference favors the picture of
stronger delocalization in 2D as compared to 1D.

Even if our data clearly show a repulsion induced delocalization effect,
they do not permit us
to draw a definite answer about the existence of a metal-insulator 
transition for this system in the thermodynamic limit.
Indeed, even though we have a significant number of fermions, we 
cannot go to larger system sizes because of the accuracy constraint
of calculating charge differences.
It should also be pointed out that our calculations done 
at $U/t \leq 2$ are below the optimal value of the interaction
strength. We expect that the delocalizing effect is even stronger 
at $U/t=4$.

In conclusion, we have studied the ground state properties of 
the repulsive Hubbard model with disorder through the powerful
PQMC method. Highly accurate simulations permit us to obtain the
difference of charge density between two ground  states with $N_p$ and
$N_p+1$ fermions respectively. The analysis of this characteristic
clearly shows that the Hubbard repulsion has a delocalizing effect
in the system. We have observed some qualitative differences between
1D and 2D for this characteristic. However, the restrictions on
system size and interaction strength do not permit us to
draw a definite conclusion about the existence of a MIT in the 
thermodynamic limit in 2D.

We thank the IDRIS at Orsay for access to their supercomputers.

\end{document}